# In Search of Future Earths: Assessing the possibility of finding Earth analogues in the later stages of their habitable lifetimes

*Running title: In Search of Future Earths*

Jack T. O'Malley-James: School of Physics and Astronomy, University of St Andrews, North Haugh, St Andrews, Fife, UK.

Jane S. Greaves: School of Physics and Astronomy, University of St Andrews, North Haugh, St Andrews, Fife, UK.

John A. Raven: Division of Plant Sciences, University of Dundee at TJHI, The James Hutton Institute, Invergowrie, Dundee, UK.

Charles S. Cockell: UK Centre for Astrobiology, School of Physics and Astronomy, James Clerk Maxwell Building, The King's Buildings, University of Edinburgh, Edinburgh, UK.

**Corresponding author:**
J.T. O'Malley-James
E.mail: jto5@st-andrews.ac.uk

Earth will become uninhabitable within the next 2-3 billion years as a result of the moving boundaries of the habitable zone caused by the increasing luminosity of the Sun. Predictions about the future of habitable conditions on Earth include a decline in species diversity and habitat extent, ocean loss and changes in the magnitudes of geochemical cycles. However, testing these predictions on the present-day Earth is difficult. The discovery of a planet that is a near analogue to the far future Earth (an old-Earth-analogue) could provide a means to test these predictions. Such a planet would need to have an Earth-like biosphere history, requiring it to have been in its system's habitable zone for giga-year (Gyr) long periods during the system's past, and to be approaching the inner-edge of the habitable zone at present. Here we assess the possibility of finding this very specific type of exoplanet and discuss the benefits of analysing older Earths in terms of improving our understanding of long-term geological and bio-geological processes. Finding such a planet in nearby star systems would be ideal, because it would be close enough to allow for atmospheric characterisation. Hence, as an illustrative example, G stars within 10 parsecs (pc) of the Sun are assessed as potential old-Earth-analogue hosts. Six of these stars are at appropriate stages of their main sequence evolution to be good potential hosts. For each of these systems, a hypothetical Earth analogue is placed at locations within the continuously habitable zone (CHZ) that would allow enough time for Earth-like biosphere development. Planetary surface temperature evolution over the host star's main sequence lifetime is assessed using a simple climate model. This is then used to determine whether the planet would be in the right stage of its late-habitable lifetime to exhibit detectable biosignatures. The best candidate in terms of the chances of planet formation in the CHZ and of biosignature detection is *61 Virginis*. Predictions from planet formation studies and biosphere evolution models suggest that only a small fraction (0.36%) of G stars in the solar neighbourhood could host an old-Earth-analogue. However, if the development of an Earth-like biosphere is assumed to be rare, requiring a sequence of low-probability events to occur, then such planets are unlikely to be found in the solar neighbourhood − although 1000s could be present in the galaxy as a whole.

**INTRODUCTION**

The fate of Earth, and of any habitable planet, is to eventually become uninhabitable as stellar evolution pushes the habitable zone boundaries further out over time (Kasting et al., 1993; Franck et al., 2000; O'Malley-James et al. 2013; Rushby et al., 2013). As an inhabited planet evolves towards this climate state, the types and abundance of life change (Kasting et al., 1993; Franck et al., 2000; O'Malley-James et al. 2013; Rushby et al., 2013). This can be predicted using climate and solar evolution models and by fitting the expected environmental conditions to the known physical and chemical limits of life. However, the means to test these predictions do not exist. Inferences about past life and climates on Earth can be supported or rejected based on geological evidence (Mojzsis et al., 1996; Nisbet & Sleep, 2001; Lepland et al., 2005). However, beyond extrapolations from the evolutionary responses of species to environmental changes (Thuiller et al., 2008), Earth provides very little evidence for predictions made about the planet's future biosphere. The only possible method for testing predictions about Earth's far future is to find and study extrasolar Earth analogues that are further along in their habitable evolution than the present-day Earth (referred to as old-Earth-analogues in this text).

On Earth, the early biosphere was composed entirely of unicellular microorganisms for approximately 2.5 Gyr, before multicellular life evolved (Butterfield, 2000; Strother et al., 2011). As the planet becomes hotter and drier as a result of the increase in solar luminosity that accompanies the later stages of the Sun's main sequence evolution, plant and animal species are expected to become extinct. This is caused by rising temperatures, which decrease water availability and increase $CO_2$ drawdown − a result of an increased atmospheric water vapour content caused by higher ocean evaporation rates. This increases precipitation rates, drawing more $CO_2$ out of the atmosphere, via silicate weathering reactions (Caldeira & Kasting, 1992). Eventually, this leaves behind an entirely microbial biosphere once again (O'Malley-James et al., 2013). This final form of the biosphere would then gradually become limited to types of life that are able to grow and reproduce under extreme conditions, until environmental conditions become too harsh even for these forms of life to survive (O'Malley-James et al, 2013; 2014). Geobiological interactions would be affected by these changes, causing atmospheric compositions and biosignatures to change over time (O'Malley-James et al., 2014).

Finding an old-Earth-analogue would enable these predictions to be tested. In this case, an old-Earth-analogue is assumed to be an Earth-mass planet in its star's habitable zone (HZ), with a similar geological and biological history to Earth. For this to be the case, the planet must have been within the continuously habitable zone (CHZ) − the region around a star that remains habitable over geological time periods − for longer than the 4.54 Gyr that Earth has spent in the solar system's CHZ. This would allow time for a complex biosphere to have evolved and for that biosphere to be declining at the present time.

To date, only one old HZ planet is known − the super-Earth Kapteyn-b (Anglada-Escudé et al., 2014). From analysis of data collected by the *Kepler* space telescope, Catanzarite & Shao (2011) estimated that Earth analogue planets are expected to exist in the HZ around 1-3% of Sun-like stars in the galaxy. More recently, Kasting et al. (2013) suggested that the frequency of HZ planets orbiting G stars is between 0.3-0.35. Using conservative estimates of the HZ (0.99-1.70 au) Petigura et al. (2013) found that up to 8.6% of Sun-like stars could host Earth-sized planets within the HZ. Approximately one third of G stars can be expected to be in the later stages of their main sequence evolution (O'Malley-James et al., 2013), making the future detection of planets in the later stages of their habitable lifetimes plausible. The proximity of any Earth-like planets found in the solar neighbourhood make them good candidates for future characterisation missions (Lunine et al., 2008). In this paper we use G stars within 10 parsecs (pc) as example targets. Stars that would be good candidate hosts of an old-Earth-analogue are identified. The chances of such a planet existing around these nearby stars, and in the galaxy as a whole, are evaluated using results from planet formation models and biosphere evolution models. The questions about Earth's future geological and biological evolution that an old-Earth-analogue could answer are discussed.

## METHODS

Stellar metallicity and stellar mass can both be related to the probability that a star hosts a planet. Marcy et al. (2005) found a strong correlation between stellar heavy metal abundance and the planet occurrence rate for gas giant planets, with more planets being found around metal-poor stars. Terrestrial planets, however, are found orbiting stars with a wide range of different metallicities (Buchhave et al., 2012). Catanzarite & Shao (2011) found that terrestrial planets orbiting Sun-like stars with an orbital semi major axis of < 1 au had the highest probability of occurrence, suggesting that G stars that are less massive than the Sun (and therefore with closer-in HZs) would be the more likely to host HZ terrestrial planets than more massive G stars. This is supported by the finding of Guo et al. (2010) who find that stars with a mass of ~0.85 $M_{\odot}$ are the most likely to host HZ planets. Using these limits as a guide, stars for this investigation are selected from the list of G stars within 10 pc of the Sun.

From the list of 20 nearby G stars, after removing those that are too young to be sufficiently evolved on the main sequence, six fit the above criteria: Alpha Centauri A (*α Cen. A*), Delta Pavonis (*δ Pav.*), Beta Hydri (*β Hyi.*), 61 Virginis (*61 Vir.*), Mu Herculis (*µ Her.*) and Beta Canum Venaticorum (*β CVn.*).

The HZ of each star system is calculated following the method of Selsis (2007) such that the inner and outer edges are defined by:

$$HZ_{inner} = (HZ_{inner,\odot} - a_{inner}\,T_* - b_{inner}\,T_*^2)\left(\frac{L_*}{L_\odot}\right)^{1/2} \qquad (1)$$

$$HZ_{outer} = (HZ_{outer,\odot} - a_{outer}\,T_* - b_{outer}\,T_*^2)\left(\frac{L_*}{L_\odot}\right)^{1/2} \qquad (2)$$

The inner and outer edges of the solar HZ are taken as 0.99 au and 1.70 au (from recent estimates - Kopparapu et al., 2013), $a_{inner} = 2.7619 \times 10^{-5}$, $b_{inner} = 3.8095 \times 10^{-9}$, $a_{outer} = 1.3786 \times 10^{-4}$, $b_{outer} = 1.4286 \times 10^{-9}$ and $T_* = T_{eff} - 5700$.

The luminosity evolution and associated effective temperature of each system over its lifetime is determined using existing models of stellar evolutionary tracks[1], using each star's estimated mass and metallicity.

The estimated age of each star is then used as a guide, within uncertainties, to determine the present extent of each star's HZ. It should be noted that the science of estimating stellar ages is not always a very precise one (Soderblom, 2010). Determining the age of single stars can be challenging, with different age determination methods resulting in a wide range of possible age values. For stars in the 0.6-1.0 $M_\odot$ range, the most precise age determination method has been found to be based on the star's rotation period - gyrochronology (Epstein & Pinsonneault, 2014). The rotation rate of low mass stars slows down in a predictable way as they age. The combination of convection and rotation in a star causes complex motions in the star's convective zone, producing and regenerating seed magnetic fields, as a result of the electrically conductive (ionised) gas in the convective zone (Maravell, 2006). Interaction between these fields and the star's ionised wind forces co-rotation of the wind to well beyond the stellar surface, causing angular momentum to be lost and resulting in a slowing of the star's rotation over time (Soderblom, 2010). Other methods, such as asteroseismology (which uses measurements of oscillation modes within a star to determine its density, and hence, age) can produce reasonably constrained age estimates, but these measurements are not always available for the stars in question and, for lower mass stars, the uncertainties associated with this method are greater (Epstein & Pinsonneault, 2014). Rotation periods can generally be accurately measured, making rotation-based age determination a reliable method for Sun-like main sequence stars. Therefore, the stellar ages used in this investigation are preferentially taken from rotation-based estimates, where possible.

---

[1] Stellar luminosity data from: http://stev.oapd.inaf.it/YZVAR/

The long-term climate evolution of Earth analogue test planets is evaluated for each system. In each case orbital semi-major axes are chosen that (i) ensure a planet will have remained within the HZ for a geologically long period of time (Gyrs) to allow for Earth-like biological evolutionary timescales, and (ii) place that planet within the boundaries of the system's HZ at the present time allowing for the possibility of present-day detectability of biosignatures.

The planetary temperature model from O'Malley-James et al. (2013) is adapted for investigating the long-term evolution of habitable conditions on Earth-like exoplanets. The incoming stellar radiation, $S_0$, is defined as

$$S_0 = \frac{L_*}{4\pi k d_p^2} \quad (3)$$

where k is a constant and $d_p$ is the mean orbital distance of the planet. The energy intercepted by the planet is then assumed to be $S_0/4$, accounting for the interception of the radiation over the Earth's spherical surface. Orbital parameter changes are accounted for to give a latitude-dependent incoming radiation value

$$S = \frac{S_0}{\pi}\left(\frac{d_p}{d}\right)^2 [h_0 \sin(\lambda)\sin(\delta) + \cos(\lambda)\cos(\delta)\sin(h_0)] \quad (4)$$

where *d* is the star-planet distance at a given point in the planet's orbit determined by the eccentricity (*e*) and true anomaly (*v*) such that

$$d = \frac{d_p(1-e^2)}{(1+e\cos(v))} \quad (5),$$

$h_0$ is the hour angle (an angular measure of the time before/after solar noon) at sunset at a given latitude, $\lambda$, and $\delta$ represents the declination of the host star in the planet's sky, depending on the obliquity of the planet ($\varphi$), the angular measure of the planet's position along its orbital path (known as the true anomaly, ν) and the longitude of perihelion (ω), i.e. a measure of the angle at which the planet makes its closest approach to the host star, such that

$$sin(\delta) = sin(\phi)sin(v + \omega) \qquad (6)$$

The obliquity and eccentricity are variable.

Some radiation is absorbed as it enters the atmosphere and does not contribute to surface heating. This absorption scales with the optical depth of the atmosphere for radiation at short wavelengths, $\tau_s$, such that the incoming radiation at the top of the atmosphere can then be defined as

$$F_{in} = (1 - a)\frac{S}{4}e^{-\tau_s} \qquad (7)$$

where $a$ is a temperature-dependent albedo. By accounting for sensible heat and latent heat losses within the atmosphere, which scale with the optical depth, and the additional warming influence of $CO_2$ and $H_2O$ within the atmosphere (the two greenhouse gases that will contribute most to the long-term temperature trends on a rapidly heating planet, as a result of ocean evaporation and the assumption of a simple relationship between $CO_2$ drawdown and increasing temperature − see later discussion), which scales with their partial pressures within the atmosphere, the balance between incoming and outgoing radiation ($F_{out} = \varepsilon\sigma T_S^4$, for surface temperature $T_S$ and emissivity $\varepsilon$) can be used to model temperature change. Following the method described in O'Malley-James et al. (2013),

$$\frac{dT}{dt} = \frac{F_{in} + F_{gh} - F_c - F_{out}}{C_p} \qquad (8)$$

where $F_{gh}$ accounts for greenhouse heating, $F_c$ accounts for convective flux within the atmosphere (sensible and latent heat) and $C_p$ is the heat capacity of the planet at constant pressure, which depends on the land-to-ocean ratio. Heat then diffuses from the equator to higher latitudes, controlled by the latitudinal diffusion coefficient as described in Lorenz *et al.* (2001).

From the surface temperature, the adiabatic lapse rate ($\Gamma_w$) can be estimated from

$$\Gamma_w = g \left[\frac{1+(H_v r/R_{sd} T)}{C_p + (H_v^2 r \varepsilon_R / R_{sd} T^2)}\right] \quad (9)$$

where $g$ is the planet's gravitational acceleration, $H_v$ is the heat of vaporisation for water, $R_{sd}$ is the specific gas constant for dry air, $\varepsilon_R$ is the ratio of the specific gas constant of dry air to that for water, $T$ is the surface temperature and $r$ is the ratio of the mass of water vapour to dry air, which depends on the saturated vapour pressure and atmospheric pressure.

More complex climate models, such as general circulation models, have been used for previous exoplanet climate studies. However, the use of a 1D energy balance model in this case lends itself to efficiently calculating surface temperatures over latitude for the ~10 Gyr time periods being investigated for each of the chosen stars. While this inevitably makes it difficult to make specific climate predictions for individual cases in this study, it does provide information on the general, long-term temperature evolution, which is convenient for comparing each of the scenarios investigated in this work.

While Earth has a near-zero eccentricity, terrestrial exoplanets generally exhibit a wide range of orbital eccentricities. The aim of this investigation is to explore the chances of detecting a very specific type of Earth-like planet: an old-Earth-analogue. Given the eccentricity range of known exoplanets, this leads to the question of how eccentric an Earth-like planet can be before annual climate variations become too extreme for the planet to have an Earth-like environmental and biological evolutionary history, rendering it too different to be a useful representative of the far future Earth. Of the discovered terrestrial planets (masses < 10 M$_\oplus$) with known eccentricities orbiting G stars, most have low eccentricities; falling within the range 0−0.1 (Schneider, 2014). This range encompasses the eccentricity variations experienced by the Earth, which varies between ~0−0.067 over 100,000 year periods. Using 0.3 as a high upper eccentricity limit, the influence eccentricity may have on long-term biosphere evolution on Earth-like planets is investigated and the suitability of high-eccentricity planets as potential old-Earth-analogues is assessed.

# RESULTS

The calculated HZ limits and stellar parameters for each of the chosen systems are presented in Table 1.

| Star system | Mass ($M_\odot$) | Age (Gyr) |
|---|---|---|
| α Cen A | 1.1 | 6.8±0.5[a] |
| δ Pav | 0.991 | 6.75±0.15[b] |
| β Hyi | 1.08 | 6.4±0.56[c] |
| 61 Vir | 0.94 | 6.35±0.25[b] |
| μ Her A | 1.0 | 6.43±0.4[d] |
| β CVn | 1.025 | 3.3-6.4[b] |

Table 1: Stellar masses and age estimates for the stars used in this study.
*a. Epstein & Pinsonneault (2014).*
*b. Mamajek & Hillenbrand (2008).*
*c. Brandão et al. (2011).*
*d. Yang & Meng (2010).*

Using equations (1) and (2) the HZ evolution for each star was calculated, as illustrated in Figure 1. The surface temperature evolution over each star's main sequence lifetime for hypothetical Earth analogues is illustrated in Figure 2. These have semi-major axes that would have been within the CHZ for geologically long periods of time in each system's past and still remain within the HZ today. In Figure 3, the influence of eccentricity on an Earth analogue planet's surface temperature evolution and habitability is presented.

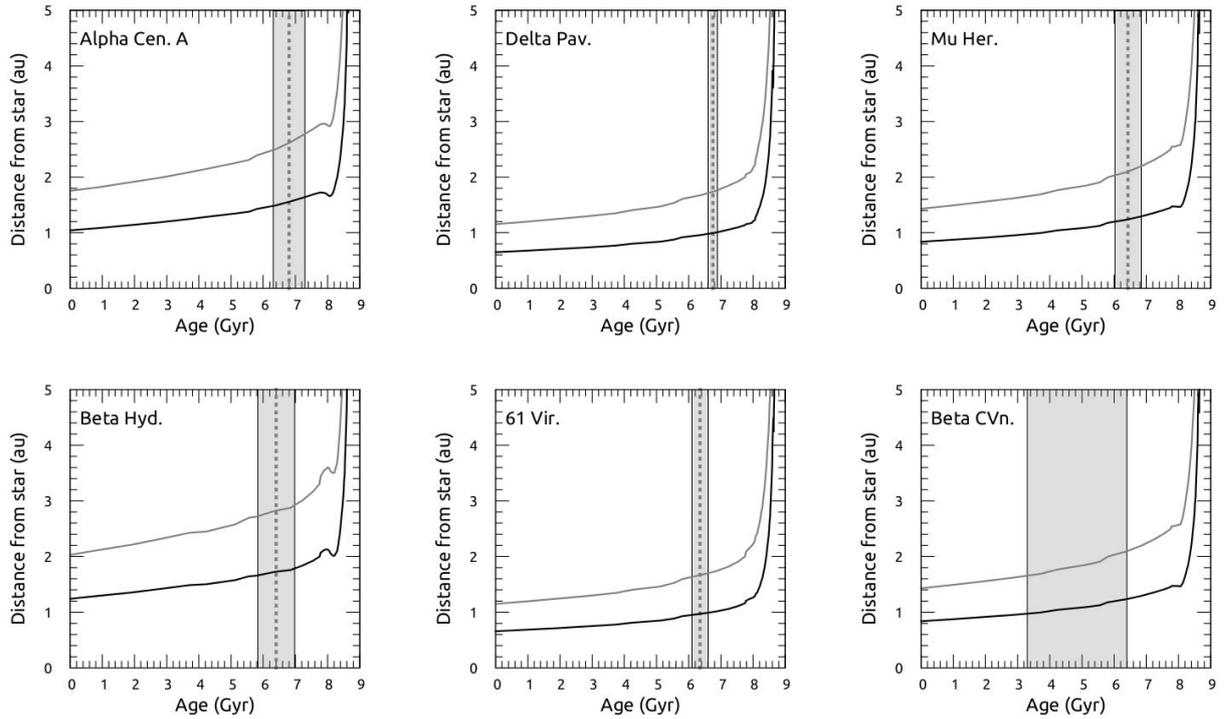

Figure 1: Habitable zone evolution during the main sequence lifetimes of the target stars. The shaded regions represent the possible ages of each star system today.

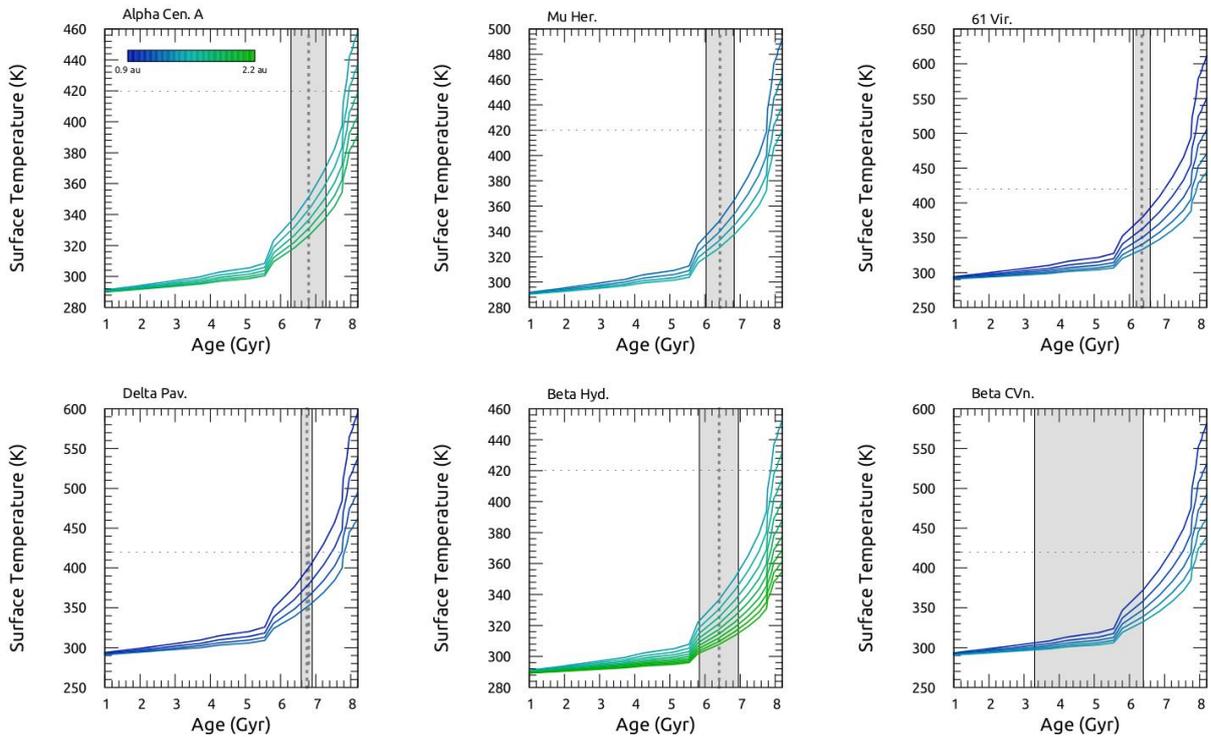

Figure 2: Temperature evolution on hypothetical Earth analogue planets orbiting each of the chosen stars. Semi-major axes are chosen that would place them at a late stage in their habitable lifetimes at the present time. The shaded regions represent the age estimates for each star system. The horizontal line at 420 K represents a theoretical upper temperature limit for life. The darker blue lines represent planets with the smallest semi-major axes. Temperatures shown are for the coolest regions on the planet's surface.

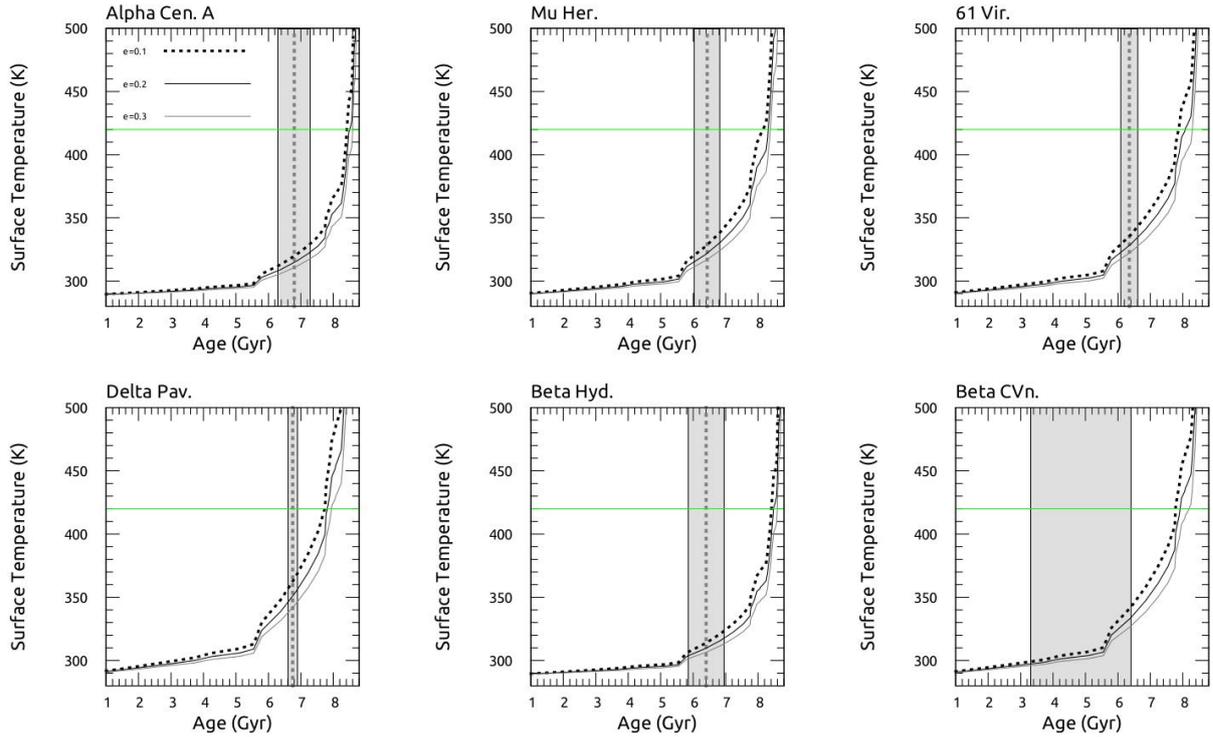

Figure 3: Temperature evolution on hypothetical Earth analogue planets with different eccentricities. In each case a semi-major axis is chosen such that the planet is in the CHZ and would have been in the CHZ for much of their star's life on the main sequence. The shaded regions represent the age estimates for each star system. The horizontal line at 420 K represents a theoretical upper temperature limit for life. Increasing eccentricity lowers mean surface temperatures; however, this temperature decrease does not alter the habitability stage the planet is expected to be in.

## DISCUSSION

### I. Finding an old-Earth-analogue

The past and predicted future habitability stages the Earth has experienced, or will experience, are summarised in Table 2. In Table 3, the systems are ranked in terms of their chances of forming planets at the chosen distances within the CHZ and are assessed in terms of the expected habitability stage of any planets orbiting at those distances. The best candidate systems from those shown in Figure 2, based on frequency estimates for terrestrial planet formation (c.f. Figure 4), are *β CVn.*, *61 Vir.* and *δ Pav.* Taking account of the ease of biosignature detectability, *61 Vir.* would be the best candidate as this could host old-Earth-analogues at the "*Microbial (declining)*" stage of their habitable lifetimes. The *δ Pav.* system would host old-Earth-analogues that would all be at very late stages in their habitable lifetimes, making biosignature detection challenging. Expected habitable conditions within the *β CVn.* system are difficult to constrain as a result of the wide range of age estimates for the star. Habitability in the other systems tends towards the "*Microbial (sparse, extremophile)*" stage, which may not produce any remotely detectable biosignatures.

| *Habitability Stage* | *Time from present (Gyr)* | *Characteristics* | *Associated biosignatures* |
|---|---|---|---|
| Pre-origin of life | -4.0 → -3.8 | Uninhabited | N/A |
| Microbial (diversifying) | -3.8 → -2.5 | Anoxic, atmospheric oxygen increasing towards the end of this stage (with the advent of oxygenic photosynthesis in unicellular organisms) | $CH_4$, Organosulphur compounds, e.g. $CH_3SH$[*] |
| Multicellular (simple) | -2.5 → -0.5 | Oxygen-rich | $O_2$, $O_3$, $N_2O$, $CH_4$, organosulphur compounds; development of vegetation red-edge[*] |
| Multicellular (diverse) | -0.5 → +0.5 | Present-day Earth-like conditions | $O_2$, $O_3$, $N_2O$, $CH_4$, $NH_3$, Organosulphur compounds; vegetation red edge |
| Multicellular (declining) | +0.5 → +1.0 | Rising temperatures; decreasing atmospheric $CO_2$ and $O_2$ levels (assuming $CO_2$ drawdown increases with temperature) | $O_2$, $O_3$, $N_2O$, $CH_4$, $NH_3$, Organosulphur compounds[**] |
| Microbial (declining) | +1.0 → +2.0 | High temperatures; anoxic | $CH_4$, Organosulphur compounds[**] |
| Microbial (sparse, extremophile) | +2.0 → ~+2.8 | High temperatures, arid, anoxic. Low levels of biological activity | $CH_4$ (weaker)[**] |

Table 2: The past and future habitability stages of Earth. Stages before the present-day period are defined by Earth's geological/geochemical past and fossil evidence for the types of life present. Future habitability stages are from the estimated timeline for the planet's far-future habitability from O'Malley-James et al. (2013).
[*] From predictions made by Pilcher (2003), Kaltenegger et al. (2007) and Seager et al. (2012).
[**] From predictions made by O'Malley-James et al. (2014).

For planets in the "*Microbial (declining)*" stage, the expected biosignatures would be similar to those for the early, pre-oxygenated Earth (cf. Table 2). However, the amount of $CH_4$ in the atmosphere during the pre-oxygenated stage on the early Earth is predicted to be 1650 ppmv (Kaltenegger et al., 2007), which is orders of magnitude higher than that predicted for the far-future Earth: ~10 ppmv (O'Malley-James et al., 2014). Hence, a far-future microbial biosphere would produce a weaker $CH_4$ signature. Low organism abundances on planets in the "*Microbial (sparse, extremophile)*" stage would make detecting any biosignatures much more challenging. Although the $CH_4$ signature would be weaker (approximately 0.1 ppmv), ocean loss would have resulted in a largely water-free atmosphere by this point, making $CH_4$ absorption bands, which would normally be obscured by water vapour, more readily detectable.

| Star system | Range of orbital semi-major axes that would allow an old Earth analogue to exist in the HZ today (au) | Expected habitable stage(s) of CHZ planets at the present time | Notes |
|---|---|---|---|
| *α Cen A* | 1.5 - 1.8 *(0.08)* | "Microbial (sparse, extremophile)" | No known planets.<br><br>Binary companion star orbiting at between 35.6-11.2 au could alter HZ boundaries, potentially making this a less viable target when searching for planets with Earth-like biosphere histories. |
| *δ Pav* | 0.9 - 1.2 *(0.15)* | An Earth analogue planet at any of the chosen orbital distances would be very close to the end of its habitable lifetime, in the "*Microbial (sparse, extremophile)*" stage, with any remaining life inhabiting small refuge environments. | The *δ Pav* system has a well constrained age estimate.<br><br>The low levels of biological activity associated with this may not produce remotely detectable biosignatures.<br><br>Radial velocity studies suggest that there are no inner-system (< 5 au) gas giant planets (Turnbull & Tarter, 2003), which could cause deviations from Earth-like habitable conditions if a rare Earth scenario is assumed. |
| *μ Her A* | 1.2 - 1.5 *(0.12)* | "*Microbial (declining)*". | No known planets.<br><br>Distant binary companion(s) |
| *β Hyd* | 1.6 - 2.2 *(0.12)* | A potential range from "*Microbial (declining)*" to "*Microbial (sparse, extremophile)*". | No known planets.<br><br>The low levels of biological activity associated with planets at the "*Microbial (sparse, extremophile)*" stage may not produce remotely detectable biosignatures. |
| *61 Vir* | 0.9 - 1.3 *(0.16)* | For any of the chosen orbital distances for the *61 Vir.* system, for any age value within its estimated age range an Earth analogue planet would be in the "*Microbial (declining)*" habitability stage. A planet at 0.9 au in this system would be closer to the "*Microbial (sparse, extremophile)*" | The case for 61 Vir. is complicated by the existence of three close-in (< 0.5 au) planets: a 5 $M_⊕$ planet, an 18 $M_⊕$ planet and an unconfirmed 23 $M_⊕$ planet (Vogt et al., 2010). Models suggest that these planets are |

| | | stage and hence, may be too near the end of its habitable lifetime to produce any detectable biosignatures, as a result of the low levels of biological activity expected on such a planet (cf. O'Malley-James et al, 2014). | in dynamically stable orbits (Vogt et al., 2010), but their effects on the long-term stability of the orbits of Earth-mass planets within the system's HZ are unknown. |
|---|---|---|---|
| β CVn | 1.0 - 1.4 *(0.20)* | The large potential age range of the *β CVn* system (between ~3.3-6.4 Gyr) means that a planet could be at habitability stages ranging from a present Earth-like stage to a "*Microbial (sparse)*" stage. | The system has a very wide range of age estimates.<br><br>No known planets. |

Table 3. An assessment of the chances of the selected nearby star systems hosting an old Earth analogue at the present time. Note: the large uncertainties associated with the age estimate of *β CVn*, make the *μ Her A* system a better candidate. The values in brackets next to the orbital semi-major axis ranges are the estimated frequency of terrestrial planets existing at this distance, based on planet formation frequency estimates summarised in Figure 4.

The results presented in Figure 3 suggest that, although increased eccentricity does influence surface temperature evolution by slowing the rate of mean temperature increase, these variations are not large enough to change the predicted habitability stage a planet would be in at the present time.

Predictions of the number of HZ planets orbiting G stars vary from the frequency of 0.3-0.35 predicted by Kasting et al. (2013), to the conservative estimates of 8.6% from Petigura et al. (2013). In Lineweaver (2001) an estimate of the average age of Earth-like planets was made. By using star formation theories as a guide for the rate of heavy metal build-up since the formation of the universe, following the argument that terrestrial planet formation correlates with heavy metal availability, it was estimated that 75% of Earth-like planets are older than Earth; on average being 1.8±0.9 Gyr older. However, requiring a planet to have an Earth-like evolutionary history further constrains the probability of the right kind of planet existing within the solar neighbourhood. Firstly, that planet would need to have been in the CHZ for geologically long periods of time, and secondly, a biosphere must have had the opportunities to develop beyond the initial microbial state (for example, without too many biosphere annihilating impact events).

*(i) Position within the CHZ:* Petigura et al. (2013) determined the occurrence rate of Earth-sized planets (radii between 1-2 $R_⊕$) as functions of orbital period. They found that 5.7% of sun-like stars could harbour an Earth-sized planet with a period of 200-400 days. In a similar study, Silburt et al. (2014) found an occurrence rate of Earth-like planets, with periods of up to 200 days and radii 1-4 $R_\oplus$, of 6.4%. However, the orbits in the case studies in this investigation are largely >400 days. Exoplanet surveys are still incomplete for the longer-period terrestrial planets that are of interest in this investigation, making an extrapolation beyond this too uncertain.

The results of planet formation simulations could provide an alternative method to estimate the occurrence frequency of longer-period planets. These model the chaotic stage of the planet formation process during which planetary embryos collide to form planets over 10-100 Myr. Typically, these simulations end with the formation of a small number of terrestrial planets in stable orbits between 0.5-2.0 au (Morbidelli et al., 2012). Raymond et al. (2004) performed 44 N-body simulations of planet formation in the solar system, taking into account a wide range of environmental conditions (including Jupiter's mass, position and eccentricity, the snow line position and the density of the solar nebula). This makes the results of these simulations particularly suitable for assessing a wide range of possible outcomes of the planet formation process. The majority of terrestrial planets (21.1%) formed between 0.6-0.8 au. For the candidate old-planet hosting systems identified here, the smallest orbital distance considered is 0.9 au. The Raymond et al. (2004) study suggests that planets are less likely to form between 0.8-1.0 au than between 1.0-1.4 au; after which, the probability of a planet existing falls off again. However, although this simulation covers a wide range of factors, it is still solar system specific.

In Raymond et al. (2011) the role of dynamic instabilities induced by giant planets on eccentric orbits was incorporated into planet formation simulations. This was motivated by the high number of eccentric giant planets that have been detected. The results of these simulations predict a lower relative frequency of terrestrial planets at all orbital distances. Planets are predicted to be more likely to occur at 0.9 au, with formation frequency decreasing with increasing orbital distance thereafter. The results of both studies are summarised in Figure 4.

Within 100 pc of the Sun, there are approximately 276 G-type stars with known age estimates, 136 of which have estimated ages of 6-10 Gyr (Holmberg et al., 2009). Based on the Petigura et al. (2013) estimate, 8.6% of these could host a HZ Earth-like planet, leaving 11 potential targets. Using the planet frequency data from Raymond et al. (2004; 2011), the mean frequency of old-Earth-analogues in these case studies is 0.14. Assuming a similar range of CHZ distances as those predicted in these case studies, 1.54 of these stars could host an old-Earth-analogue. Hence, one old-Earth-analogue could exist within 100 pc (0.36% of the G stars in the solar neighbourhood). Assuming this holds for all G stars in the galaxy (approximately $1.5 \times 10^{10}$ stars), then approximately $5 \times 10^7$ could host an old-Earth-analogue.

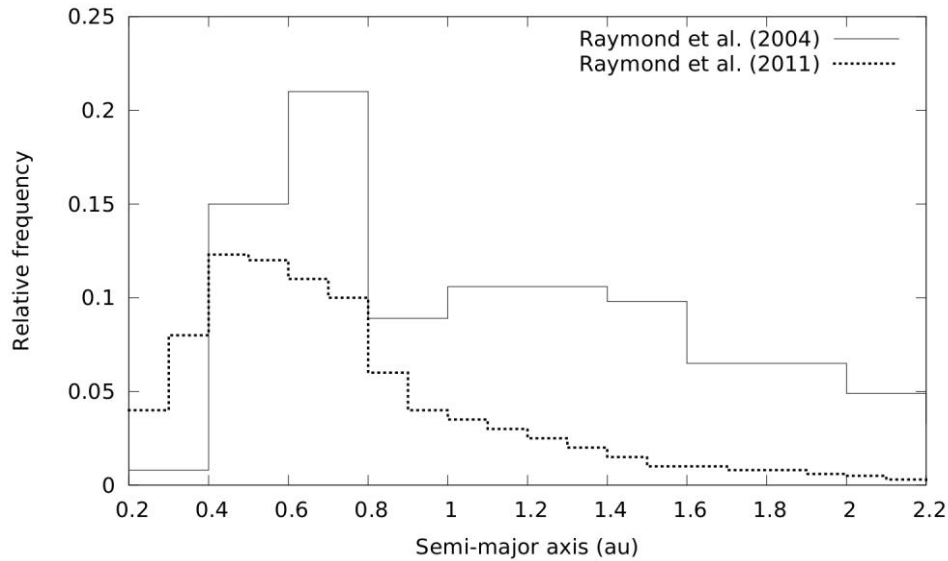

Figure 4: Summary of the frequency of planet formation at a given orbital distance − from simulations by Raymond et al. (2004) and Raymond et al. (2011).

Table 4 summarises other stages in the habitable evolution of Earth-like biospheres and the distances at which an Earth-like planet would have to orbit, within each of the case study star systems, to currently be at one of these stages. Based on the results of the planet formation studies summarised in Figure 4, the expected frequencies of planets with *Earth-like biospheres* and *early microbial biospheres* are an order of magnitude lower than those for *old-Earth-analogue biospheres*. In all cases, the expected frequency of previously habitable planets (those that have already reached the ends of their habitable lifetimes) is approximately double the expected frequency of old-Earth-analogues. The expected frequencies of planets that have never been habitable are the highest in most cases, with the exceptions of *δ Pav*. and *61 Vir*. These stars have the lowest masses out of the six stars in the case studies, resulting in closer-in HZs. As planet formation frequencies are higher closer to the star, this makes HZ planets more likely around low-mass G stars. This suggests that the best candidate systems for finding old-Earth-analogues are low-mass G stars in the later stages of their main sequence evolution. However, for older G stars, more previously habitable planets that no longer host a biosphere should be expected. These will likely have existed closer to the inner-edge of their system's HZ and will not have remained within the HZ for long enough to have undergone the expected future changes to habitable conditions on Earth. Hence, looking at planets orbiting younger stars would not be a successful strategy for finding an analogue to Earth's future habitability. In the wake of recent HZ planet discoveries and expectations of future discoveries, these results highlight the need to consider the temporal nature of habitability alongside a planet's current position within a star's HZ.

| Star | HZ Distances at which different biospheres could exist (au) | | | |
|---|---|---|---|---|
| | *Early Microbial Biosphere* | *Earth-like Biosphere* | *Extinct Biosphere* | *Never Habitable* |
| α Cen A | 2.1 - 2.4 *(0.02)* | 1.8 - 2.1 *(0.06)* | 1.0 - 1.5 *(0.22)* | < 1.0 *(0.57)* |
| δ Pav | 1.4 - 1.5 *(0.05)* | 1.2 - 1.4 *(0.12)* | 0.6 - 0.9 *(0.30)* | < 0.6 *(0.26)* |
| μ Her A | 1.7 - 1.9 *(0.04)* | 1.5 - 1.7 *(0.04)* | 0.8 - 1.2 *(0.25)* | < 0.8 *(0.47)* |
| β Hyd | 2.4 - 2.6 *( - )* | 2.2 - 2.4 *(0.02)* | 1.2 - 1.6 *(0.17)* | < 1.2 *(0.65)* |
| 61 Vir | 1.4 - 1.5 *(0.05)* | 1.3 - 1.4 *(0.06)* | 0.6 - 0.9 *(0.30)* | < 0.6 *(0.26)* |
| β CVn | 1.8 - 1.9 *(0.04)* | 1.4 - 1.8 *(0.13)* | 0.8 - 1.0 *(0.20)* | < 0.8 *(0.47)* |

Table 4. The HZ distances at which different biospheres could exist in each of the case study star systems. *Early Microbial* biospheres resemble life on the early Earth and will have spent up to 2 Gyr in the HZ. *Earth-like* biospheres resemble the present day biosphere on Earth. *Extinct* biospheres account for planets that were in the HZ in the system's past, but are now inwards of the inner-edge of the HZ. *Never Habitable* planets are those that have always been inwards of the inner-edge of the HZ for the host star's lifetime. The values in brackets next to the orbital semi-major axis ranges are the estimated frequency of terrestrial planets existing at this distance, based on planet formation frequency estimates summarised in Figure 4.

*(ii) Biosphere development:* Events such as large asteroid strikes can delay or prevent the evolution of a simple biosphere into a more complex one. The *61 Vir.* system, for example, is known to have a larger debris disc than the solar system, hosting ten times more comets (Wyatt et al., 2012). This could result in a higher impact frequency for any inner planets, stalling biosphere development. Using Monte Carlo simulations to determine the number of stars with habitable planets, Forgan & Rice (2010) estimate the fraction of those planets that could host a complex biosphere like that on Earth, rather than being either lifeless or stuck in a microbial stage of evolution as a result of sterilising events, such as impacts or nearby supernovae. If simply being in the CHZ is enough for a complex biosphere to develop, they estimate that $10^7$ stars (out of a total of $10^9$) in simulated galaxies could host an Earth-like habitable planet. However, if a rare Earth scenario is assumed, in which factors such as a sun-like star, an Earth-like planet mass, a large nearby moon and a solar system formation history are required, that estimate drops to $1\times10^3$; $1\times10^{-4}$ % of stars in the galaxy. This would suggest that the solar neighbourhood is likely to be devoid of habitable planets with advanced biospheres.

## *II. What outstanding questions about future planetary evolution on Earth could be answered by finding an old-Earth-analogue?*

*The temperature-silicate weathering connection:* Silicate weathering results in the drawdown of $CO_2$ from the atmosphere via precipitation reactions with silicate rocks. The products of this reaction dissolve in soil water and ultimately reach the oceans (Raven & Edwards, 2001). Here, ocean chemistry results in the deposition of half of the carbon drawn down from the atmosphere over timescales of up to 10s Myr (Berner & Berner, 1996; Raven & Edwards, 2001). The $CO_2$ is eventually released back into the atmosphere via outgassing from the mantle; a process that takes 100s Myr. Any $CaCO_3$ that reaches the land is weathered, such that over 10-100 Myr timescales, $CO_2$ drawdown and release balance (Raven & Edwards, 2001).

Silicate weathering rates have been linked to surface temperature, with a higher surface temperature leading to a higher $CO_2$ drawdown rate (West et al., 2005). This has led to claims that this acts as a climate stabiliser, removing more of the greenhouse gas $CO_2$ as temperatures increase. However, silicate weathering depends on a number of other factors, with the rate of the dissolution reaction depending on factors including the availability of water, silicate substrates, relief and acidity (predominantly controlled by $CO_2$ availability, but also by organic acids produced by vegetation) (West et al., 2005; Tyrrell, 2014). While there is evidence that it is accelerated by warmer conditions, there is also observational evidence that this is not always the case. Other factors, such as relief and rock type can play more important roles in some cases (Tyrrell, 2014).

Long-term future climate predictions for Earth are based on the $CO_2$ drawdown rate increasing in response to increasing temperature, until atmospheric $CO_2$ levels are depleted to levels too low for photosynthesis to take place. Given that there are other factors that can influence the silicate weathering rate, these predictions may turn out to be over-simplified. Observations of $CO_2$ (or a lack thereof) in the atmosphere of an old-Earth-analogue, could help to justify or refute this simplifying assumption, i.e. more $CO_2$ than expected could imply that $CO_2$ drawdown does not simply increase with temperature.

*2. Mantle cooling − connection to the rate of tectonic activity and volcanism:* The rate of mantle degassing is known to be linked to the vigour of mantle convection (Padhi et al., 2012). Until recently, it was assumed that the cooling of the mantle over geological time has slowed mantle convection, slowing tectonic activity and outgassing rates. This leads to the assumption that in Earth's far future, the core will have cooled to such an extent that tectonic activity eventually stops. However, evidence suggests that tectonic activity was actually slower on the early Earth, despite a hotter core (Korenaga, 2013). This may be a result of a hotter core causing deeper melting in the mantle, causing the mantle to become more viscous, slowing convection (Korenaga, 2013).

Another long-held belief about tectonics is that water facilitates plate movements. The presence of small amounts of water can weaken rocks and minerals; a process called hydrolytic weakening. Regenauer-Lieb et al. (2001) claimed that this weakening of rocks in the presence of water enables the initiation of the subduction of plates. This leads to the conclusion that in the far future, as surface temperatures increase and the planet starts to lose water, plate movements may stop, regardless of the core-temperature−mantle-viscosity relation (Meadows 2007, O'Malley-James et al., 2013). However, recent work by Fei et al. (2013) suggests that the hydrolytic weakening effect in olivine is far less than previously thought. This may mean that the effect of water on facilitating plate movements is not as great as assumed, potentially contesting far-future predictions of the rate of tectonic activity.

Modelling efforts and continued study of the mechanisms behind plate tectonics could help constrain the predictions about the rate of tectonic activity based on core temperature and water availability. However, if exoplanet tectonic activity could be inferred, predictions about tectonic activity on Earth could be improved, i.e. if a greater than predicted rate of tectonic activity was inferred on an old-Earth-analogue, this could provide evidence against the predicted decline in tectonic activity on the far-future Earth. At present, determining rates of tectonic activity on exoplanets would be very difficult (Spiegel et al., 2013); however, it may be achievable through measurements of volcanic gas levels in a planet's atmosphere (probably $SO_2$; cf. Kaltenegger et al. (2010)) from which volcanic activity could be estimated, enabling some inferences about tectonic activity to be made. Although it should be noted that atmospheric $SO_2$ can also be produced by oxidation of biogenic $(CH_3)_2S$.

*3. Ocean loss:* One aspect of the predicted future sequence of events leading to the end of Earth's habitable lifetime is increased ocean evaporation as temperatures rise. The roles of different factors, such as cloud cover and atmospheric dynamics, in potentially enhancing or delaying this process, are only beginning to be understood (Leconte et al., 2013). An example of a planet entering this runaway ocean evaporation stage could help support and refine modelling efforts.

**CONCLUSIONS**

The future of life on Earth is linked to the future main sequence evolution of the Sun, which is expected to alter the planetary environment by raising temperatures and driving runaway heating and the gradual extinction of the biosphere. These predictions can only be verified by finding old-Earth-analogue planets. The temperature models in this work suggest that suitable host stars exist within 10 pc of the Sun. However, these planets are probably rare. Estimates in this study suggest that only one such planet would exist within 100 pc − a distance that places it close enough for atmospheric characterisation to be feasible. If conservative criteria on biosphere evolution are imposed, this number falls to only ~$10^3$ in the galaxy as a whole. The search for old-Earth-analogues is also hampered by the difficulties associated with determining a star's age, leading to uncertainties in estimates of the habitable stage a planet may have reached. Better gyrochronology data is needed to improve age estimates of G stars, which would enable old-Earth-analogues to be identified with more confidence.

**ACKNOWLEDGEMENTS**

This research has made use of the VizieR catalogue access tool, CDS, Strasbourg, France". The original description of the VizieR service was published in A&AS 143, 23 (2000). The authors acknowledge the comments from an anonymous reviewer that helped to shape our arguments. The University of Dundee is a registered Scottish charity, No SC015096. JTO acknowledges an STFC Aurora grant.

**Author disclosure statement**

No competing financial interests exist.